\documentclass[twocolumn,nofootinbib,floatfix,superscriptaddress]{revtex4}        
\usepackage{graphicx}
\usepackage{epsfig}
\usepackage{bm}
\usepackage[T1]{fontenc}
\usepackage[latin9]{inputenc}
\usepackage{amssymb}
\usepackage{float}
\usepackage{amsmath}
\usepackage{dcolumn}
\usepackage{cancel}
\usepackage[colorlinks]{hyperref}
\usepackage[usenames,dvipsnames]{color}
\hypersetup{
	breaklinks=true,
	pdfstartview={FitH},    
	colorlinks=true,       
	linkcolor=blue,          
	citecolor=red,        
	filecolor=magenta,      
	urlcolor=blue,           
	anchorcolor=green,      
	linktocpage=true
}

\newcommand{\be}{\begin{equation}}
\newcommand{\ee}{\end{equation}}

\begin{document}

\title{Universal thermodynamic topological classes of static black holes in Conformal Killing Gravity}
\author{Hao Chen}
\email{haochen1249@yeah.net}
\affiliation{School of Physics and Electronic Science, Zunyi Normal University, Zunyi 563006,PR China}
\author{Di Wu}
\email{wdcwnu@163.com}
\affiliation{School of Physics and Astronomy, China West Normal University, Nanchong, Sichuan 637002, People's Republic of China}
\author{Meng-Yao Zhang}
\email{myzhang94@yeah.net}
\affiliation{College of Computer and Information Engineering, Guizhou University of Commerce, Guiyang, 550014, China}
\author{Soroush Zare}
\email{szare@uva.es}
\affiliation{Departamento de F\'{i}sica Te\'{o}rica, At\'{o}mica y Optica and Laboratory for Disruptive Interdisciplinary Science (LaDIS),
Universidad de Valladolid, 47011 Valladolid, Spain}
\author{Hassan Hassanabadi}
\email{hha1349@gmail.com}
\affiliation{Departamento de F\'{i}sica Te\'{o}rica, At\'{o}mica y Optica and Laboratory for Disruptive Interdisciplinary Science (LaDIS),
Universidad de Valladolid, 47011 Valladolid, Spain}
\affiliation{Department of Physics, University of Hradec Kr\'{a}lov\'{e}, Rokitansk\'{e}ho 62, 500 03 Hradec Kr\'{a}lov\'{e}, Czechia }
\author{Bekir Can L\"{u}tf\"{u}o\u{g}lu}
\email{bekircanlutfuoglu@gmail.com}
\affiliation{Department of Physics, University of Hradec Kr\'{a}lov\'{e}, Rokitansk\'{e}ho 62, 500 03 Hradec Kr\'{a}lov\'{e}, Czechia }
\author{Zheng-Wen Long}
\email{zwlong@gzu.edu.cn}
\affiliation{College of Physics, Guizhou University, Guiyang, Guizhou 550025, People's Republic of China}

\begin{abstract}
In this study, we develop universal thermodynamic topological classes for the static black holes in the context of the Conformal Killing Gravity. Our findings indicate that the Conformal Killing Gravity significantly reconstructs the thermodynamic properties of  both the smallest inner and the largest outer black hole states. Additionally, it considerably alters the thermodynamic stability of black holes across both high-temperature and low-temperature  regimes. This analysis shows that different CKG parameter settings will lead to $W^{0+}$ $(\lambda>0)$ and $W^{1+}$ $(\lambda<0)$  categories for the charged AdS black hole, the Reissner-Nordstr$\ddot{o}$m black hole in Conformal Killing Gravity is classified into the $W^{0+}$ and $W^{1+}$ categories. Furthermore, we examine the specific scenario where charge is neglected. The study reveals that within the framework of Conformal Killing Gravity, the Schwarzschild black hole similar to the Schwarzschild-AdS black hole, can be classified into the $W^{1-}$ and $W^{0-}$ categories. This work provides key insights into the fundamental nature of quantum gravity theory.
\end{abstract}

\maketitle

\section{Introduction}
The groundwork for black hole thermodynamics was established by considering black holes emitting thermal radiation as conventional thermodynamic systems. Exploring the interplay between the essential characteristics of black holes and thermodynamic principles acts as a link that connects classical gravity, quantum mechanics, and thermodynamics \cite{ch0}. Hawking and Page initially employed the black hole as a thermodynamic system, revealing that the thermal stability of the black hole is governed by the critical temperature, it is named as the Hawking-Page phase transition \cite{ch01}. Subsequently, the cosmological constant was reinterpreted as a thermodynamic pressure \cite{ch4,ch5,ch6,ch7,ls0,ls1}, which stimulated in-depth investigations into different kinds of phase transitions \cite{myy1,myy2,myy3,myy4,myy5,myy6,myy7,myy8,myy9,myy10,jyy1,jyy2} and Joule-Thomson effects \cite{myy11,myy12,myy13,myy14,myy15,myy16,myy17,myy18} in black holes. As a crucial analytical approach, thermodynamic topology provides valuable supplementary perspectives for understanding black hole thermodynamics. Wei et al. proposed two novel methodologies: First, by constructing a thermodynamic function of temperature, critical points segregate into two classes (conventional or novel) based on their topological indices \cite{G2}. Second, the black hole solution is envisioned  as the topological defect, the generalized off-shell free energy is employed to analyze the topological classification (three classes) of black hole thermodynamics \cite{G1}. These methods have been widely adopted to analyze the  different kinds of black holes \cite{ll1,gx2,ll2,ll3,ll4,ll5,ll6,ll7,ll8,ll9,ll34,ll10,ll11,ll12,ll13,ll14,ll15,ll16,ll17,ll18,ll19,
ll20,ll21,ll22,ll23,ll24,ll25,ll26,ll27,yin1,ll28,ll29,ll30,ll31,ll32,ll33}.

Next, we will offer a brief explanation of the generalized off-shell Helmholtz free energy $(\mathcal{F})$ concept and present an overview of the core elements of the topological method. To this end, we can identify a black hole characterized by its entropy and mass  within a cavity \cite{G1}, $\mathcal{F}$ can be represented as follows:
\begin{equation}
\label{n1}
\mathcal{F}=M-\frac{S}{\tau},
\end{equation}
under the condition that the  inverse temperature parameter $(\tau)$  meets the condition
$\tau=\beta=1 / T $, as discussed in \cite{gx1}, the equation (\ref{n1}) is reduced to
\begin{equation}
F=M-TS.
\end{equation}
To conduct a comprehensive examination of the topological characteristics, we introduce a supplementary parameter $\Theta$ taking values in the interval $(0, \pi)$.  This enables the establishment of a two-part vector field, which may be represented as \cite{G1}
\begin{equation}\label{GS1}
\phi=\left(\phi^{r_h}, \phi^{\Theta}\right)=\left(\frac{\partial \tilde{\mathcal{F}}}{\partial r_h}, \frac{\partial \tilde{\mathcal{F}}}{\partial \Theta}\right),
\end{equation}
here, the function $\tilde{\mathcal{F}}$ is defined as
\begin{equation}
\tilde{\mathcal{F}}=\mathcal{F}+\frac{1}{\sin \Theta}.
\end{equation}
Notably, the condition $\phi^{r_{h}}=0$, characterizes the black hole state as the zero point associated with the vector field. In accordance with the  theory of mapping topological current, this is strongly connected to a topological charge \cite{kh6}, which is obtained by the winding number $w$. If the winding number is positive, this suggests that the black hole is in a stable condition, with the associated heat capacity exceeding zero. Conversely, if the winding number is negative, it implies that the black hole is in an unstable state, and the corresponding heat capacity is less than zero. The overall sum of all winding numbers is referred to as the index $W=\sum_{i=1}^N w_i$, which establishes the theoretical foundation for the topological categorization of black holes and provides a cohesive interpretation of their thermal stability and phase characteristics.

Recently, in the study of black hole topology, Wei et al. proposed a pioneering method. By conducting a thorough examination of the asymptotic characteristics of the inversion temperature parameter, black hole solutions have been divided into four categories $(W^{1+},W^{0-},W^{0+},W^{1-}$) \cite{gx1}. In this regard, Wu et al. conducted a more systematic expanded of the universal topological classification, and developed a new topological classification along with two additional topological subcategories \cite{gx4}. Up to now, this approach has also been utilized to explore various types of black holes, including three-dimensional rotating Ba$\tilde{n}$ados-Teitelboim-Zanelli black holes \cite{gx3}, higher-dimensional rotating Kerr black holes \cite{gx5}. Additionally, it has been applied to analyze in the dark matter background \cite{gx6}. It is crucial to emphasize that Conformal Killing Gravity (CKG), as a novel gravitational correction framework, not only retains all solutions of general relativity but also effectively overcomes the intrinsic theoretical constraints of general relativity, thereby expanding the range of large-scale gravitational phenomenology studies. Within this context, undertaking a thorough examination of the universal thermodynamic topology of  black holes carries substantial theoretical significance and constitutes the primary motivation for this research exploration.

The structure of this paper is arranged as follows: In Sec.\ref{II}, a brief overview is presented regarding the characteristics and  thermodynamic quantities the of charged AdS black holes in CKG background. In Sec.\ref{III}, we focus on the thermodynamic topological classification of static black holes, with detailed analyses of both universal properties and specific cases.  The research findings will be summarized and discussed in Sec.\ref{IIII}.
\section{Black holes in CKG background and thermodynamic quantities} \label{II}
General relativity elucidates the intrinsic relationship between space-time and gravity by interpreting gravity as the curvature of space-time geometry, thereby significantly enhancing humanity's comprehension of the structure of space-time and the nature of gravity. The detection of gravitational waves generated by the collision of two black holes serves as definitive proof of black holes' existence \cite{my1,my2}. Following this, the visual representations of the supermassive black hole \cite{my3,my4,my5,my6} and Sagittarius A* \cite{my7,my8}, which is situated at the center of the Milky Way, have been disclosed. The dark areas encircled by bright light show striking agreement with the theoretical expectations of black hole shadows as outlined by general relativity. Although general relativity is effective in explaining gravitational phenomena at local and solar scales, it encounters significant challenges when applied to galactic and cosmic scales, such as its inability to account for the galactic dynamics and the universe's accelerated expansion \cite{my9,my10}. To tackle these contradictions, the standard cosmological model incorporates the concepts of dark energy and dark matter \cite{my11,my12}, along with alternative gravity theories \cite{my13}, which are challenging to verify through experiments. Recently, the Cotton gravity theory has gained attention as an extension of general relativity. This interest is largely due to its association with several foundational problems within general relativity \cite{my14,my15,mm2.1,mm2,mm1}. This special theory can enhance  our comprehension of the vacuum structure. On this basis, Harada further developed a new gravitational expansion \cite{mm3,mm3.1,mm4}, and the corresponding field equation is expressed as follows:
\begin{equation}\label{ch1}
H_{\mu \nu \rho}=8 \pi G T_{\mu \nu \rho},
\end{equation}
here, $H_{\mu \nu \rho}$  represents the Cotton tensor, and exhibits complete symmetry in the indices $\mu$, $\nu$, and $\rho$, and fulfills the traceless condition expressed as $g^{\nu \rho} H_{\mu \nu \rho}=0$. While $T_{\mu \nu \rho}$ is associated with the energy-momentum tensor, and entirely symmetric and adheres to the relation $g^{\nu \rho} T_{\mu \nu \rho}=2 \nabla_\nu T_\mu^\nu $, which subsequently implies the conservation law $\nabla_\nu T_\rho^\nu=0$. These are explicitly given by \cite{HaradaPRD2023}
\begin{equation}
\begin{aligned}
H_{\mu \nu \rho}& \equiv  \nabla_\rho \mathcal{R}_{\mu \nu}+\nabla_\mu \mathcal{R}_{\nu \rho}+\nabla_\nu \mathcal{R}_{\rho \mu} \\
& -\frac{1}{3}\left(g_{\mu \nu} \partial_\rho+g_{\nu \rho} \partial_\mu+g_{\rho \mu} \partial_\nu\right) \mathcal{R},
\end{aligned}
\end{equation}
\begin{equation}
\begin{aligned}
T_{\mu \nu \rho}& \equiv  \nabla_\rho T_{\mu \nu}+\nabla_\mu T_{\nu \rho}+\nabla_\nu T_{\rho \mu} \\
& -\frac{1}{6}\left(g_{\mu \nu} \partial_\rho+g_{\nu \rho} \partial_\mu+g_{\rho \mu} \partial_\nu\right) T,
\end{aligned}
\end{equation}
where, $\mathcal{R}$ represents the Ricci tensor, while $T$ stands for the conventional energy-momentum tensor. Within the framework of Cotton gravity, the field equation can be represented in a parameterized form, which is analogous to the field equations, which have been altered through the incorporation of a divergence-free conformal Killing tensor \cite{my13}, specifically,
\begin{equation}
R_{\nu \rho}-\frac{1}{2} R g_{\nu \rho}=T_{\nu \rho}+K_{\nu \rho},
\end{equation}
\begin{equation}
\begin{aligned}
&\left( \nabla_\mu K_{\nu \rho}+\nabla_\nu K_{\mu \rho}+\nabla_\rho K_{\mu \nu}\right) \\
&=\frac{1}{6}\left(g_{\nu \rho} \nabla_\mu K+g_{\mu \rho} \nabla_\nu K+g_{\mu \nu} \nabla_\rho K\right).
\end{aligned}
\end{equation}
Building on this, the general static spherically symmetric metric reads
\begin{equation}
\label{metric2}
ds^2 = -f(r)\,dt^2 + \frac{dr^2}{f(r)} + r^2\,d\theta^2 + r^2 \sin^2\theta\,d\phi^2.
\end{equation}
In the context of CKG theory, one can derive the solution for the charged AdS black hole \cite{my15,mm5}, where its metric function can be represented as
\begin{equation}\label{f2}
f(r) = 1 - \frac{2M}{r} + \frac{q^2}{r^2} - \frac{\Lambda}{3}r^2 - \frac{\lambda}{5}r^4.
\end{equation}
In this context, $M$ denotes the mass, $\Lambda$ stands for the cosmological constant, $q$ indicates the existence of a nonzero electric charge, and $\lambda$ refers to the Conformal Killing Gravity. We can observe that the CKG takes precedence when the $r$ region approaches infinity in Eq. (\ref{f2}). Additionally, the parameter $\lambda$ must remain sufficiently small in absolute value to maintain consistency with general relativity at low energy \cite{my15}. If the nonlinear charge $(q=0)$ is ignored, the black hole becomes a Schwarzschild-AdS type in the CKG background \cite{mm3}.

In order to investigate the thermodynamic properties described in \eqref{metric2}, it is beneficial to adopt the extended phase space framework. In this regard, the corresponding thermodynamic quantities \cite{mk1} are represented as follows:
\begin{equation}
\label{c1}
S = \pi r_h^2,
\end{equation}
\begin{equation}\label{c2}
M(r_h)=\frac{r_h}{2}+\frac{q^2}{2 r_h}+\frac{4}{3} P \pi r_h^3-\frac{\lambda r_h^5}{10},
\end{equation}
\begin{equation}
T\left(r_h\right)=\frac{1}{4 \pi r_h}-\frac{q^2}{4 \pi r_h^3}+2 P r_h-\frac{\lambda r_h^3}{4 \pi}.
\end{equation}

\section{Topological classes of static black holes in CKG background }\label{III}
In this section, our main goal is to investigate the universal thermodynamic topological categories of static black holes within the CKG framework. By substituting the entropy (as given in Eq. (\ref{c1})) and the Hawking temperature (from Eq. (\ref{c2})) into Eq. (\ref{n1}), we can obtain
\begin{equation}
\mathcal{F}=\frac{r_h}{2}-\frac{\pi r_h^2}{\tau}+\frac{q^2}{2 r_h}+\frac{4}{3} P \pi r_h^3-\frac{\lambda r_h^5}{10}.
\end{equation}
The off-shell free energy enables us to introduce a novel function that includes an extra parameter $(\Theta)$, which is expressed as
\begin{equation}
\tilde{\mathcal{F}}=\frac{r_h}{2}-\frac{\pi r_h^2}{\tau}+\frac{q^2}{2 r_h}+\frac{4}{3} P \pi r_h^3-\frac{\lambda r_h^5}{10}+\frac{1}{\sin \Theta},
\end{equation}
the components associated with the charged-AdS black hole in the CKG background are presented as
 \begin{equation}
\phi^{r_h}=\frac{1}{2}\left(1-\frac{q^2}{r_h^2}-\frac{4 \pi r_h}{\tau}+8 \pi P r_h^2-\lambda r_h^4\right)
\end{equation}
and
\begin{equation}
\phi^{\Theta}=-\csc \Theta \cot \Theta .
\end{equation}
By taking into account the condition $\phi^{r_h}=0$, we are able to determine
 \begin{equation}
\tau=\beta=\frac{4 \pi r_h^3}{r_h^2-q^2+8 P \pi r_h^4-\lambda r_h^6}.
\end{equation}
Next, we shall explore the universal thermodynamic topological classification under varying black hole parameter conditions.\\
\begin{figure}[t]
\begin{center}
\includegraphics[width=0.4\textwidth]{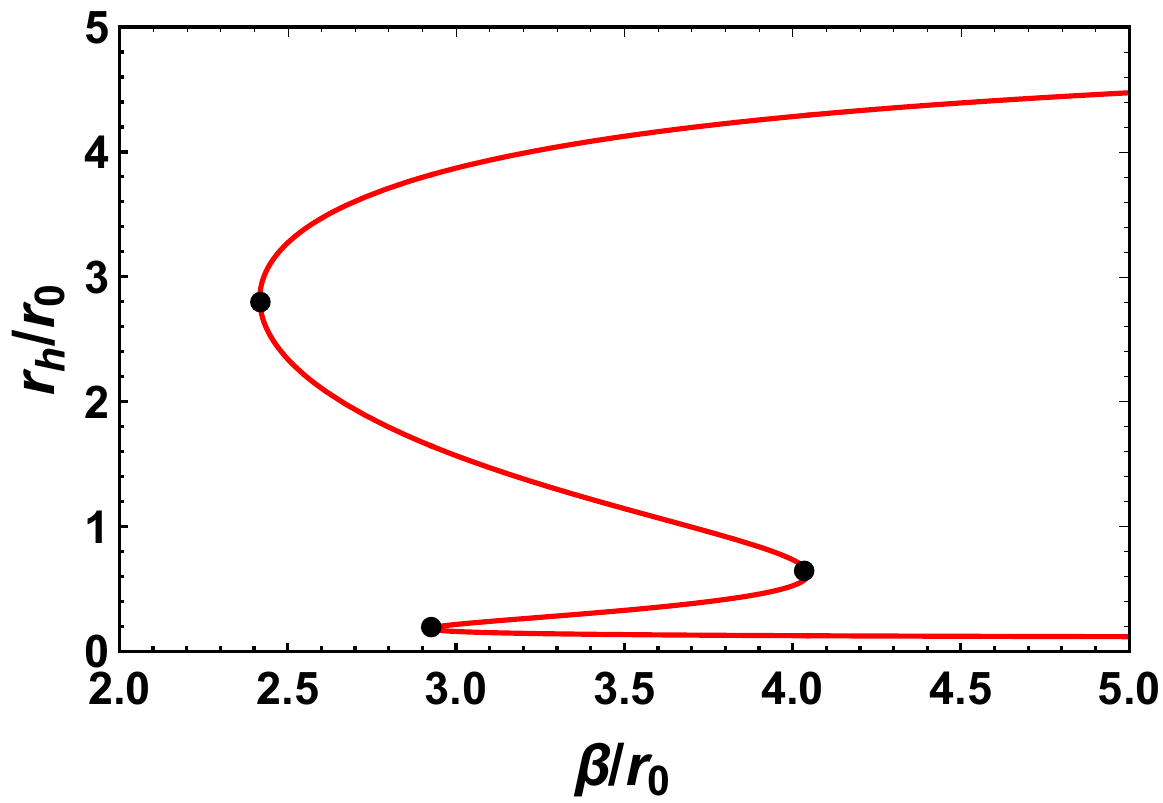}
\caption{The vector field $\phi^{r_{h}}$ is depicted in the $r_{h} / r_0$-$\beta / r_0$ plane, where $P r_0^2=0.1$,$q / r_0=0.1 $ and $\lambda / r_0=0.1$.}
\label{meng1}	
\end{center}
\end{figure}
\subsection{The positive CKG parameter}
We first consider cases where the CKG parameter is positive, while constraining the inversion temperature parameter to
\begin{equation}\label{jj1}
\lambda>0, \quad \beta\left(r_m\right)=\infty\quad \text { and }\quad \beta(\infty)=\infty.
\end{equation}
We proceed to study the asymptotic characteristics of $\phi$ in the vicinity of the boundary described by Eq. (\ref{jj1}), this boundary is represented by the closed contour $C = I_1 \cup I_2 \cup I_3 \cup I_4$, where each segment $I_i$ is expressed as follows:
\begin{equation}
\begin{aligned}
&I_1 = \{r_h = \infty, \Theta \in (0, \pi)\}, \\
&I_2 = \{r_h \in (\infty, r_m), \Theta = \pi\},\\
&I_3 = \{r_h = r_m, \Theta \in (\pi, 0)\},\\
&I_4 = \{r_h \in (r_m, \infty), \Theta = 0\}.
\end{aligned}
\end{equation}
\begin{figure}[t]
\begin{center}
\includegraphics[width=0.45\textwidth]{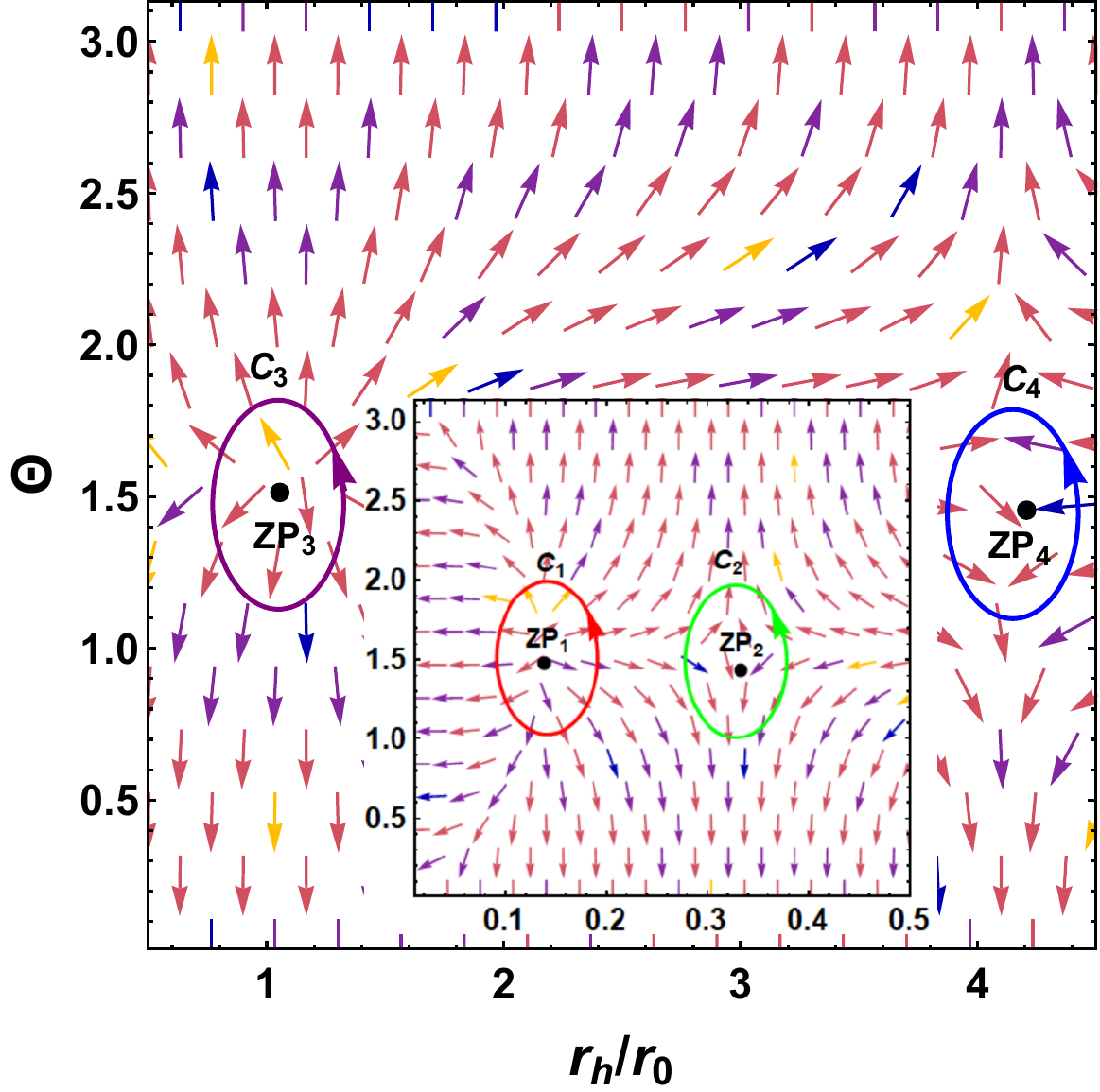}
\caption{The arrow signifies that, for the unit vector field of the black hole on the $r_{h}-\Theta$ plane  for $P r_0^2=0.1$, $q / r_0=0.1$, $\tau / r_0=3.5$ and $\lambda / r_0=0.1$, the zeros labeled as $ZP_{1}$,$ZP_{2}$, $ZP_{3}$ and $ZP_{4}$ with black dots.}
\label{meng2}	
\end{center}
\end{figure}
The contour $C$ encompasses the complete parameter space of relevance. Importantly, the design of $\phi$ ensures its orthogonality to the segments $I_2$ and $I_4$ \cite{G1}, which indicates that the primary asymptotic  behavior of $\phi$ are determined along $I_1$ and $I_3$. As $r_h$ approaches either $r_m$ or infinity, The vector $\phi$ demonstrates a shift directed towards the left, with its angle determined by the component value \(\phi^\Theta\) .
To visually explore the connection between the inversion temperature parameter and the horizon radius, we refer to \cite{mm8,mm9} for the exact values of parameters employed in this study.
The relationship between $r_{h} / r_0$ and $\beta/ r_0$ is depicted in FIG.(\ref{meng1}), with $P r_0^2$, $q / r_0$, and $\lambda / r_0$ held constant at $0.1$. Importantly, the annihilation/generation points appear at specific $\beta_{c}$ values and must meet the following conditions:
\begin{equation} \label{po1}
\frac{\partial \beta}{\partial r_h}=\frac{\partial^2 \beta}{\partial r_h^2}=0.
\end{equation}
If we ignore the influence of charge $(q=0)$, we can obtain that the annihilation/generation points is determined by
\begin{equation}
\beta_c=\frac{3 \sqrt{3} \pi \sqrt{\frac{4 P \pi \mp \sqrt{16 P^2 \pi^2-3 \lambda}}{\lambda}} \lambda}{16 P^2 \pi^2 \mp 4 P \pi \sqrt{16 P^2 \pi^2-3 \lambda}+3 \lambda}.
\end{equation}
Given the constraint in  Eq. (\ref{po1}) and the parameter assignment $P r_0^2=0.1$,$q / r_0=0.1 $ and $\lambda / r_0=0.1$, we can find two generation points $(\beta_{c1}/ r_0=2.4171, \beta_{c2}/ r_0=2.9273)$ and one annihilation point $(\beta_{c3}/ r_0=4.0467,)$ in FIG.(\ref{meng1}).
\begin{figure}[ht!]
\begin{center}
\includegraphics[width=0.4\textwidth]{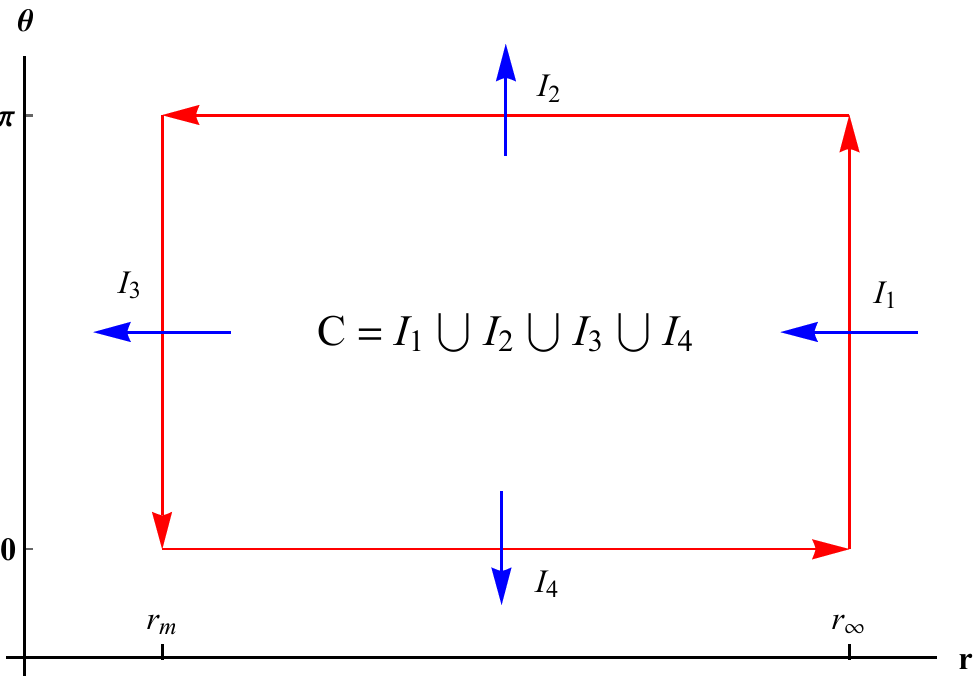}
\caption{The asymptotic behavior of vector field $\phi$ at at the boundary $(C=I_1 \cup I_2 \cup I_3 \cup I_4$), the blue arrows indicate the direction of the vector field for
 the charged-AdS black hole in CKG background $(\lambda>0)$.}
\label{fig:Ic1}	
\end{center}
\end{figure}
\begin{figure}[t]
\begin{center}
\includegraphics[width=0.4\textwidth]{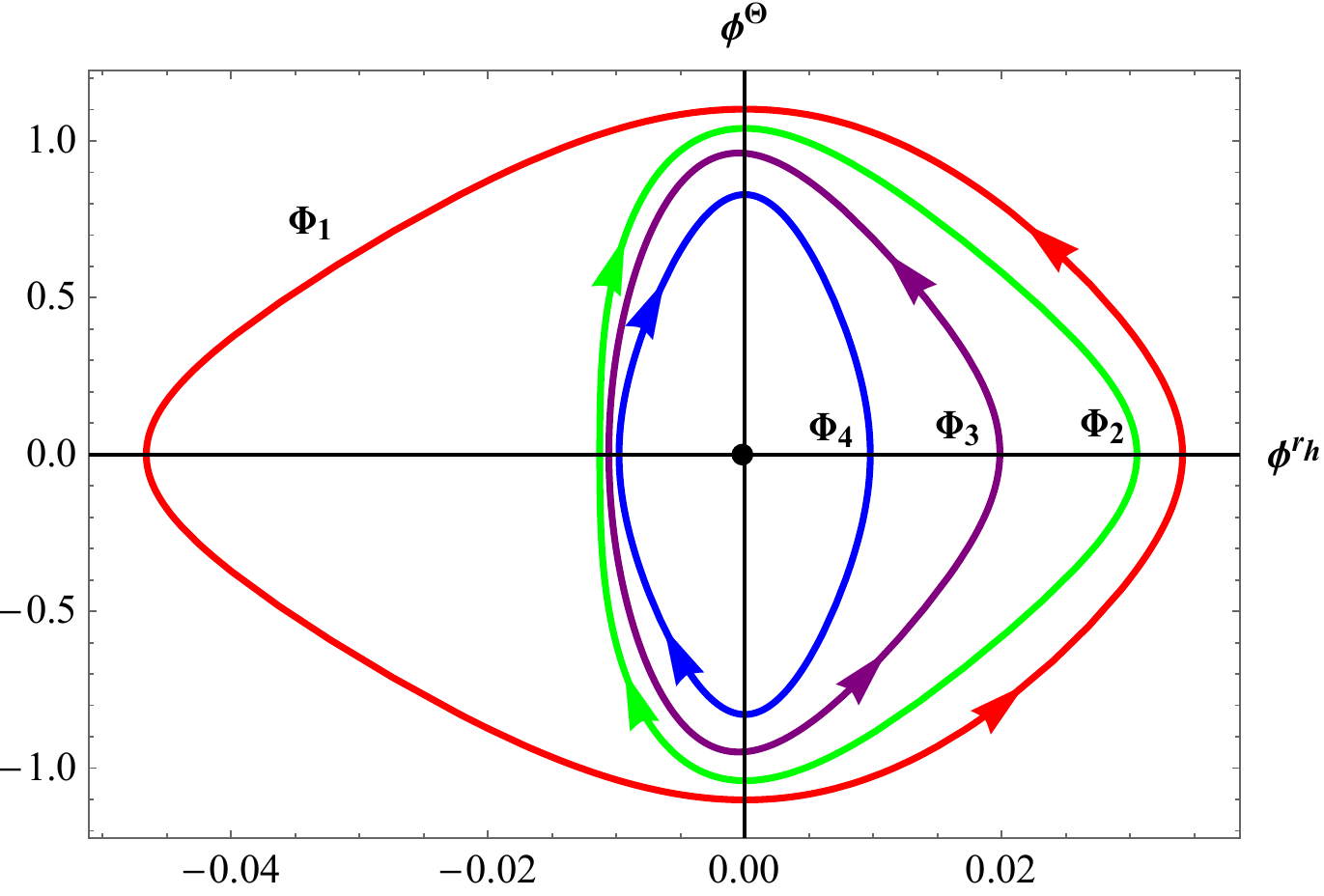}
\caption{The contours $\Phi_i$ illustrate the variations in the components of the vector field $\phi$ as the paths $C_i$ depicted in FIG.(\ref{meng2}) are followed for the charged-AdS black hole within the CKG background. Zero points are labeled with black dots, the winding numbers of $\Phi_{2}$ and $\Phi_{4}$ are -1, whereas those of $\Phi_{1}$ and $\Phi_{3}$ are 1.}
\label{meng3}	
\end{center}
\end{figure}
The vector field, corresponding to the charged-AdS black hole within the CKG framework, is illustrated in FIG.(\ref{meng2}). The asymptotic properties of this vector field near the boundaries $I_i$ are summarized in Table(\ref{table3}). It can be observed that as $r_h$ approaches $r_m$, or when $r$ tends toward infinity, the vector field plot of $\phi$ exhibits a leftward orientation, indicative of its universal thermodynamic classification.
In Fig. (\ref{fig:Ic1}), the contour $C$ is plotted to show the asymptotic behavior of the vector field $\phi$ for the charged-AdS black hole $\lambda>0$.

The changes in the components for $\Phi_i$ can be visualized through the contours $C_i$ presented in FIG.(\ref{meng3}). For $\lambda = 0.1$, the winding numbers $\Phi_1$ and $\Phi_3$ equal $1$, while $\Phi_2$ and $\Phi_4$ equal $-1$. Noted that the thermal stability of the system in the low temperature and the high temperature regions is different. More specifically, when $\beta \rightarrow \infty$, the system includes a stable small black hole alongside an unstable large black hole. On the contrary, under the condition $\beta \rightarrow 0$, the existence of a black hole state is not possible. Consequently, according to the universal thermodynamic class introduced in Ref. \cite{gx1}, the charged-AdS black hole under the CKG framework $(\lambda>0)$ falls into the category of $W^{0+}$.\\

\subsection{The negative CKG parameter}
If we assume that the CKG parameter is negative, the asymptotic behavior of the parameter $\beta$ can be described by the following limits:
\begin{equation}
\lambda < 0, \quad \beta\left(r_m\right)=\infty \quad \text { and } \quad \beta(\infty)=0.
\end{equation}
\begin{figure}[t]
\begin{center}
\includegraphics[width=0.4\textwidth]{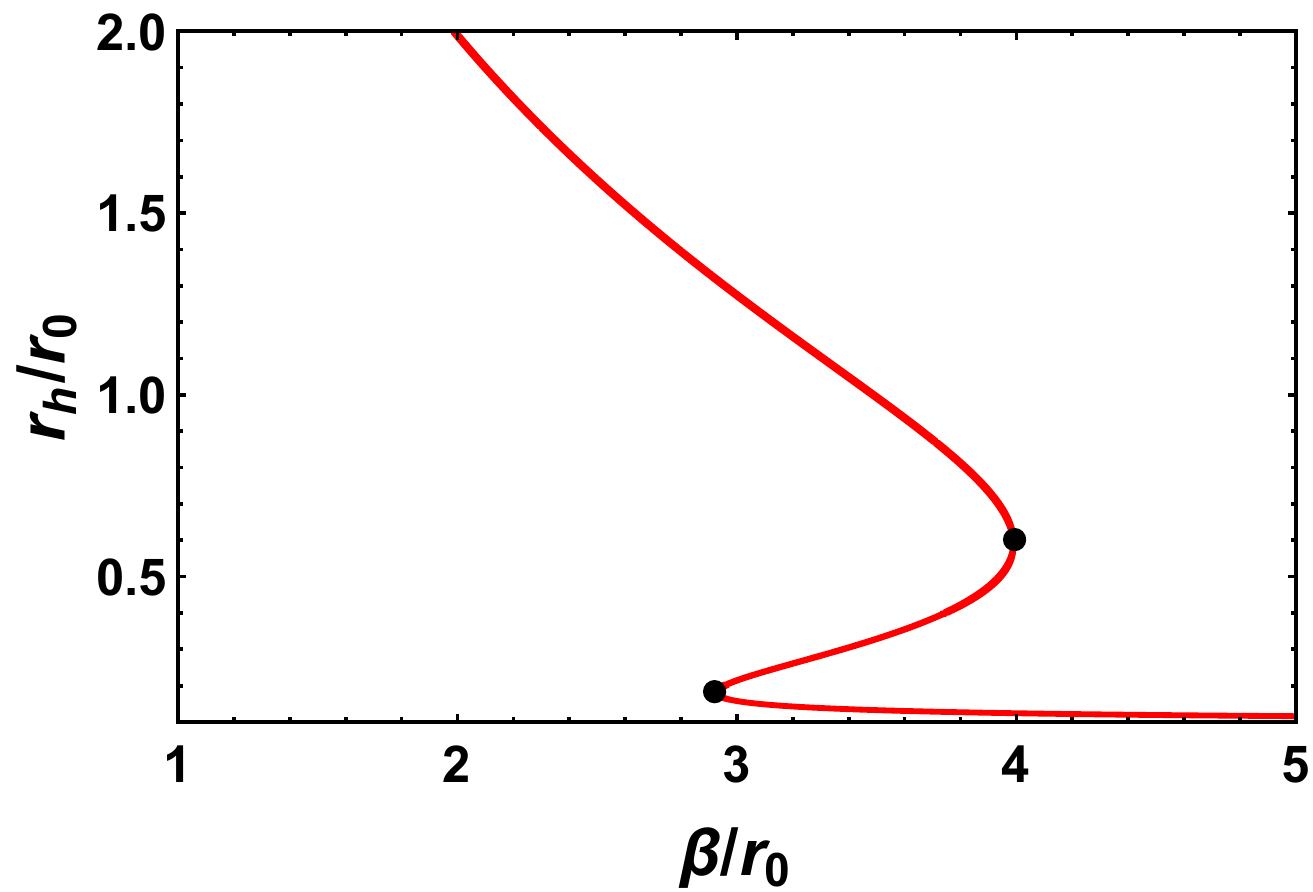}
\caption{The vector field $\phi^{r_{h}}$ is depicted in the $r_{h} / r_0$-$\beta / r_0$ plane, where $P r_0^2=0.1$, $q / r_0=0.1 $ and $\lambda / r_0=-0.1$.}
\label{meng4}	
\end{center}
\end{figure}
\begin{figure}[t]
\begin{center}
\includegraphics[width=0.3\textwidth]{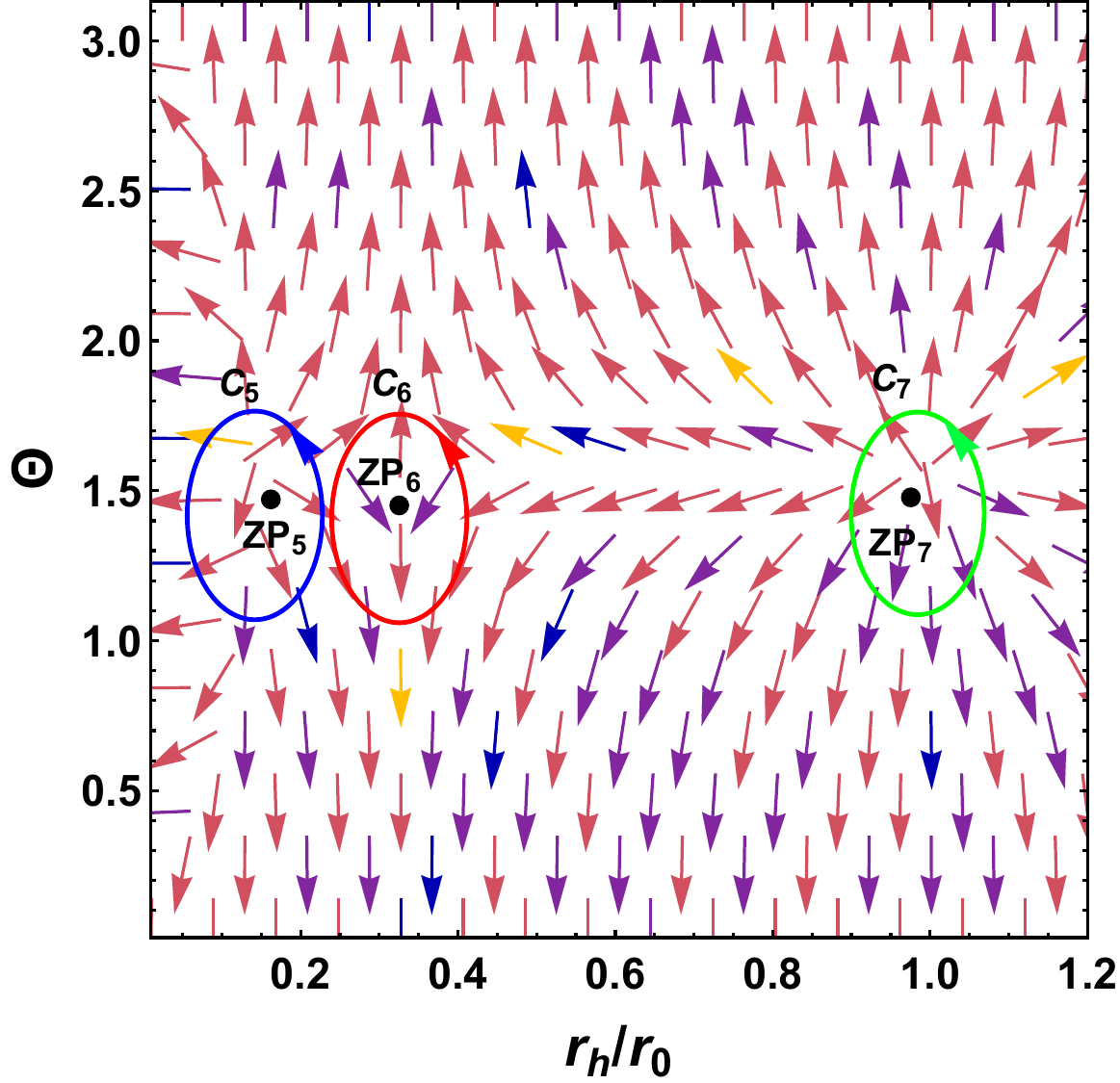}
\caption{The arrow signifies that, for the unit vector field of the black hole on the $r_{h}-\Theta$ plane  for $P r_0^2=0.1$, $q / r_0=0.1$, $\tau / r_0=3.5$ and $\lambda / r_0=-0.1$, the zeros labeled as $ZP_{5}$,$ZP_{6}$ and $ZP_{7}$ with black dots.}
\label{meng5}	
\end{center}
\end{figure}
\begin{figure}[t]
\begin{center}
\includegraphics[width=0.4\textwidth]{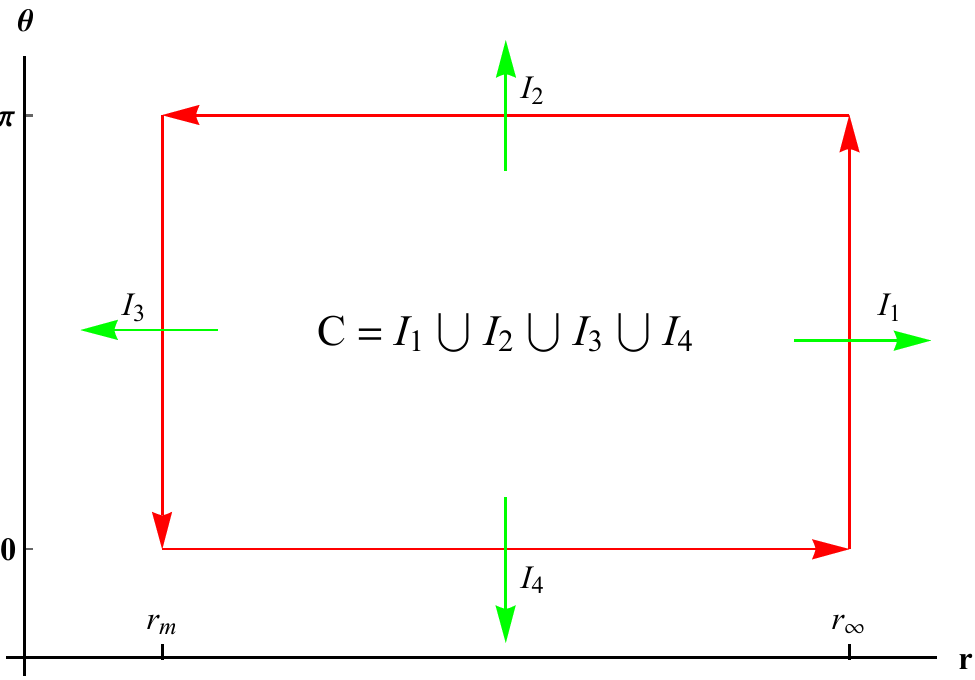}
\caption{The asymptotic behavior of vector field $\phi$ at at the boundary $(C=I_1 \cup I_2 \cup I_3 \cup I_4$), the green arrows indicate the direction of the vector field for
 the charged-AdS black hole in CKG background $(\lambda<0)$.}
\label{fig:Ic2}	
\end{center}
\end{figure}
\begin{figure}[t]
\begin{center}
\includegraphics[width=0.4\textwidth]{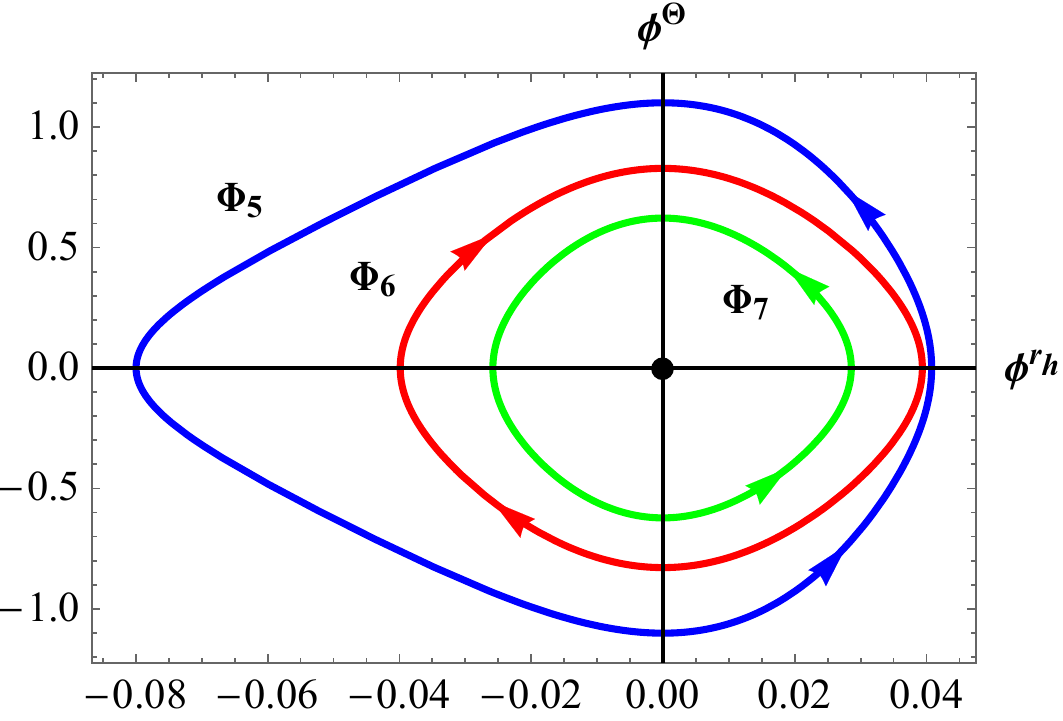}
\caption{The contours $\Phi_i$ illustrate the variations in the components of the vector field $\phi$ as the paths $C_i$ depicted in FIG.(\ref{meng5}) are followed for the charged-AdS black hole within the CKG background. Zero points are labeled with black dots, the winding numbers of $\Phi_{6}$ is -1, whereas those of $\Phi_{5}$ and $\Phi_{7}$ are 1.}
\label{meng6}	
\end{center}
\end{figure}
It is evident that the CKG parameter has a significant influence on both the black hole generation point and the annihilation point. More specifically, when $\lambda$ is set to $\lambda = -0.1$, we can find that there is a generation point at $\beta_{c1}/ r_0=2.9265$ and an annihilation point at $\beta_{c2}/ r_0=3.9898$, as illustrated in FIG.(\ref{meng4}). Furthermore, it is noteworthy that the vector field direction at boundary $I_1$ points to the right, while at boundary $I_3$, it points to the left, as summarized in Table (\ref{table3}). The vector field associated with the charged-AdS black
hole under the CKG framework is illustrated in FIG.(\ref{meng5}).
In Fig.(\ref{fig:Ic2}), the contour $C$ is plotted to show the asymptotic behavior of the vector field $\phi$ for the charged-AdS black hole $\lambda <0$.

The directional changes of $\phi$ across the contours are clearly depicted in FIG.(\ref{meng6}). The winding numbers $\Phi_{5}$ and $\Phi_{7}$ are both equal to $1$, while $\Phi_{6}$ equals $-1$. Large and small black holes demonstrate a heat capacity that exceeds zero, which suggests their stability. By contrast, intermediate-sized black holes  possess a heat capacity that is below zero, rendering them unstable. In the low-temperature regime as $\beta \rightarrow \infty$, the system exhibits stable black holes of small size and unstable black holes of intermediate size. Conversely, in the high-temperature region as $\beta \rightarrow 0$, intermediate-sized black holes that are unstable can be found coexisting with large-sized black holes that remain stable. Consequently, the charged-AdS black hole within the framework of CKG background $(\lambda<0)$ falls into the thermodynamic topological class $W^{1+}$. When the effects of CKG are disregarded $(\lambda=0)$, the topological classification $(W^{1+})$ of the charged-AdS black hole aligns perfectly with the findings reported in \cite{gx3}.\\
\subsection{The Schwarzschild-AdS black hole in the CKG background}
We analyze the Schwarzschild-AdS black hole in the CKG background, wherein the asymptotic characteristics of the parameter $\beta$ satisfy the following constraints:
\begin{equation}
\begin{aligned}
&\lambda>0, \quad \beta\left(r_m\right)=0, \quad \text { and } \quad \beta(\infty)=\infty,\\
&\lambda < 0, \quad \beta\left(r_m\right)=0, \quad \text { and } \quad \beta(\infty)=0.
\end{aligned}
\end{equation}

\begin{figure}[t]
\begin{center}
\includegraphics[width=0.4\textwidth]{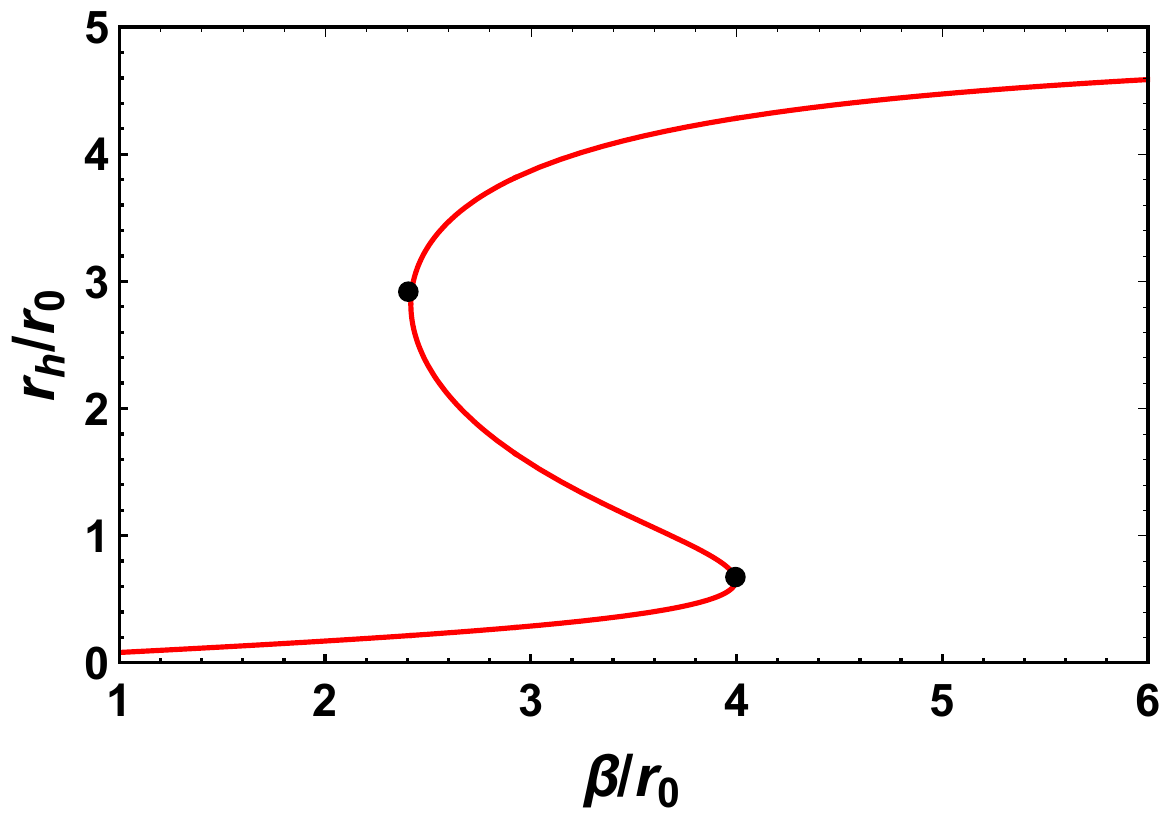}
\caption{The vector field $\phi^{r_{h}}$ is depicted in the $r_{h} / r_0$-$\beta / r_0$ plane, where $P r_0^2=0.1$ and $\lambda / r_0=0.1$.}
\label{zl1}	
\end{center}
\end{figure}
\begin{figure}[t]
\begin{center}
\includegraphics[width=0.4\textwidth]{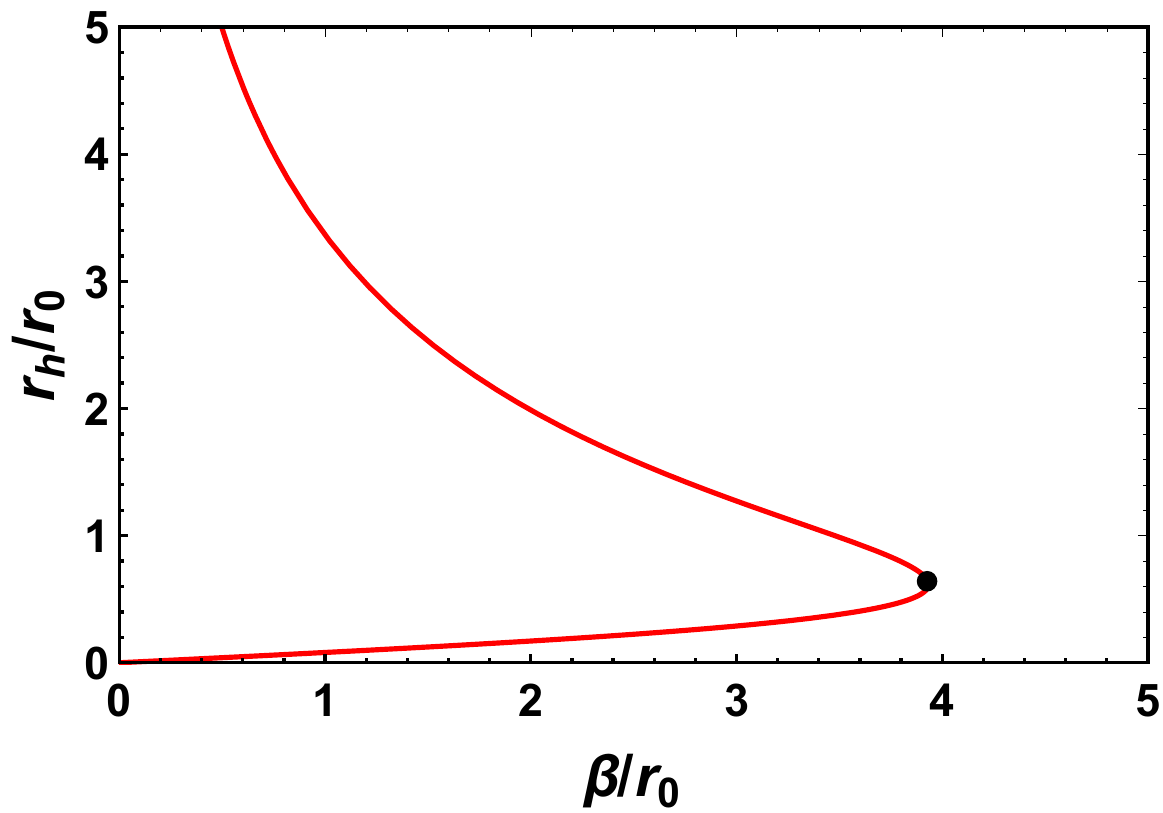}
\caption{The vector field $\phi^{r_{h}}$ is depicted in the $r_{h} / r_0$-$\beta / r_0$ plane, where $P r_0^2=0.1$ and $\lambda / r_0=-0.1$.}
\label{zl2}	
\end{center}
\end{figure}
\begin{figure}[t]
\begin{center}
\includegraphics[width=0.3\textwidth]{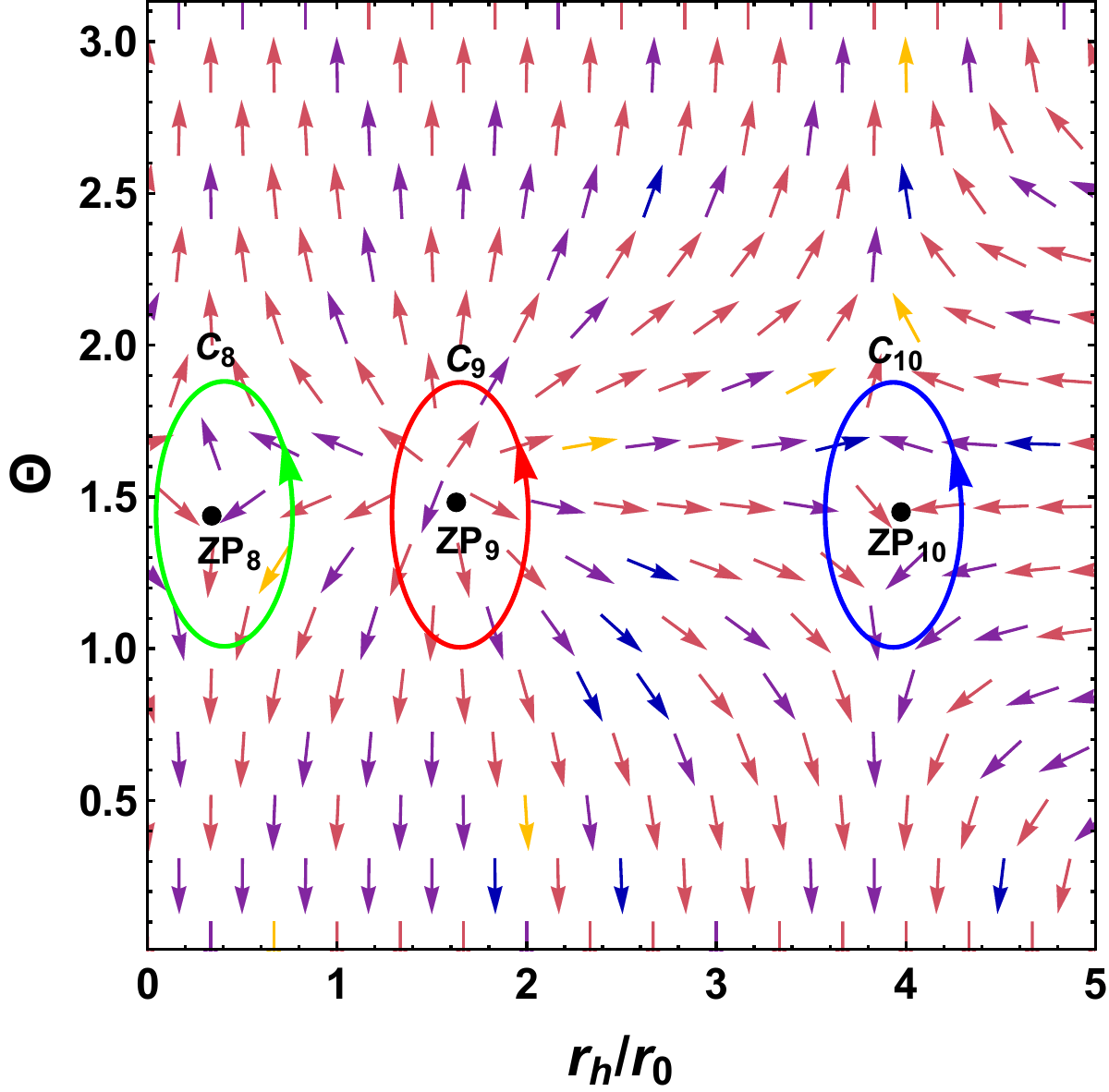}
\caption{ The arrow signifies that, for the unit vector field of the black hole on the $r_{h}-\Theta$ plane  for $P r_0^2=0.1$, $\tau / r_0=3$ and $\lambda / r_0=0.1$, the zeros labeled as $ZP_{8}$,$ZP_{9}$ and $ZP_{10}$ with black dots.}
\label{zl3}	
\end{center}
\end{figure}
\begin{figure}[t]
\begin{center}
\includegraphics[width=0.3\textwidth]{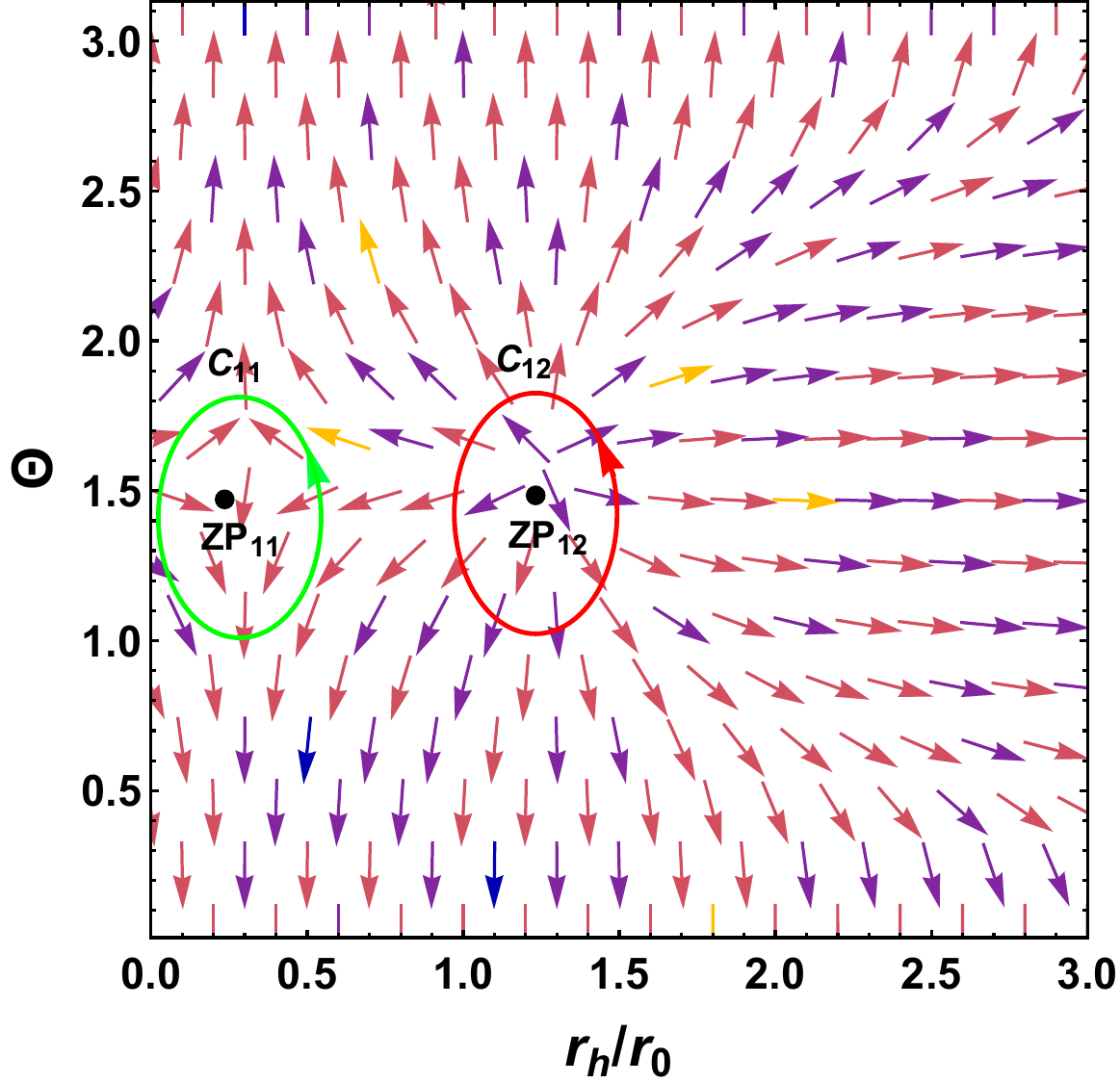}
\caption{ The arrow signifies that, for the unit vector field of the black hole on the $r_{h}-\Theta$ plane  for $P r_0^2=0.1$, $\tau / r_0=3$ and $\lambda / r_0=-0.1$, the zeros labeled as $ZP_{11}$ and $ZP_{12}$ with black dots.}
\label{zl4}	
\end{center}
\end{figure}
\begin{table*}[ht]
    \caption{The orientation of the black hole in CKG background is indicated by the arrow of $\phi^{r_{h} }$, which is associated with the corresponding topological number.}
    \centering
    \begin{tabular}{|l|c|c|c|c|c|}
        \hline
        \textbf{Black hole solutions }& ${I_1}$ & $I_2$ & ${I_3}$ & ${I_4}$ & ${W}$ \\
        \hline
        the charged-AdS black hole in CKG background $(\lambda>0)$ & $\leftarrow$ & $\uparrow$ & $\leftarrow$ & $\downarrow$ & 0 \\
        \hline
        the charged-AdS black hole in CKG background $(\lambda<0)$& $\rightarrow$ & $\uparrow$ & $\leftarrow$ & $\downarrow$ & 1 \\
        \hline
        Schwarzschild-AdS black hole in  CKG background $(\lambda>0)$ & $\leftarrow$ & $\uparrow$ & $\rightarrow$ & $\downarrow$ & -1 \\
        \hline
        Schwarzschild-AdS black hole in  CKG background $(\lambda<0)$& $\rightarrow$ & $\uparrow$ & $\rightarrow$ & $\downarrow$ & 0 \\
        \hline
    \end{tabular}
   \label{table3}
\end{table*}
\begin{figure}[t]
\begin{center}
\includegraphics[width=0.4\textwidth]{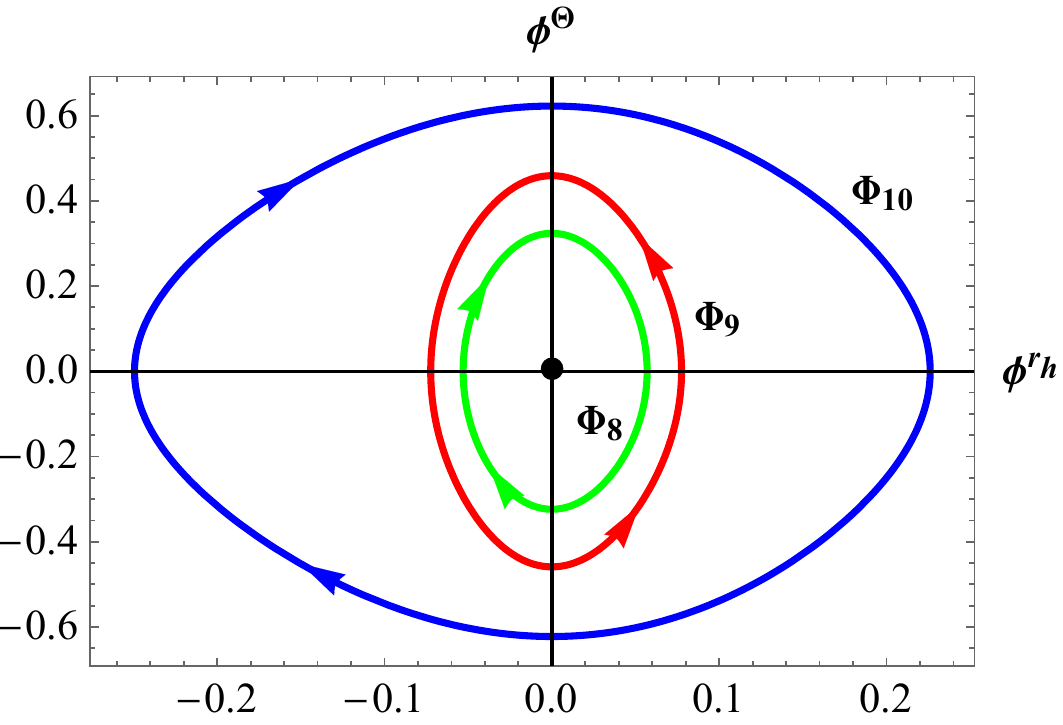}
\caption{The contours $\Phi_i$ illustrate the variations in the components of the vector field $\phi$ as the paths $C_i$ depicted in FIG.(\ref{zl3}) are followed for the Schwarzschild-AdS black hole in CKG background . Zero points are labeled with black dots, the winding numbers of $\Phi_{8}$ and $\Phi_{10}$ are -1, whereas those of $\Phi_{9}$ is 1.}
\label{zl5}	
\end{center}
\end{figure}
\begin{figure}[t]
\begin{center}
\includegraphics[width=0.4\textwidth]{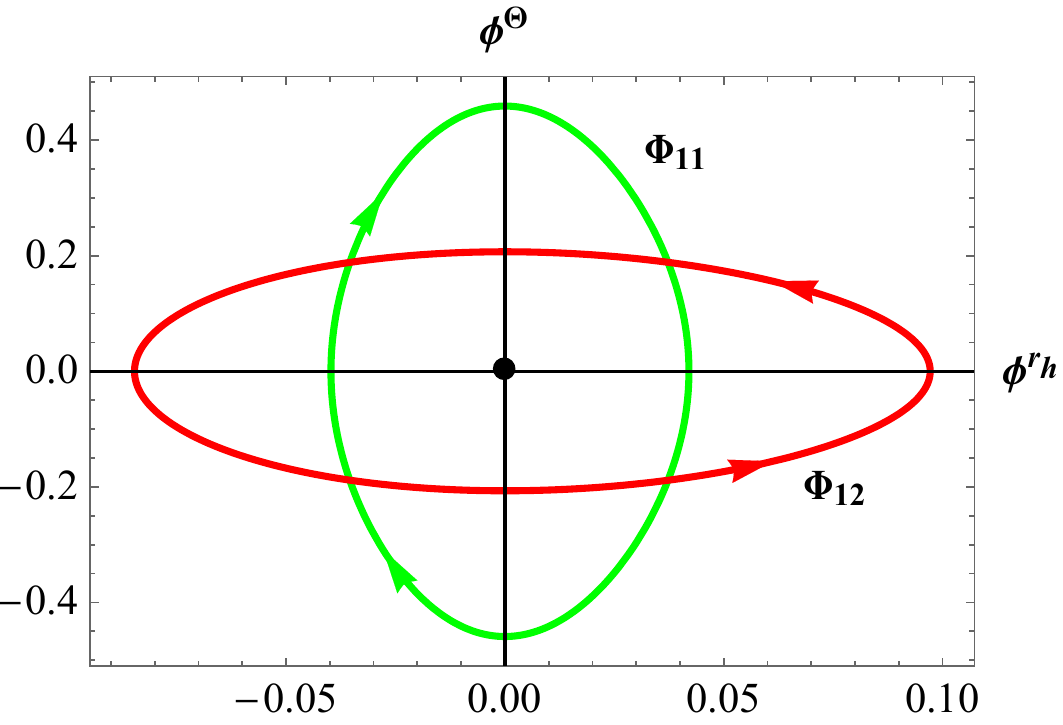}
\caption{The contours $\Phi_i$ illustrate the variations in the components of the vector field $\phi$ as the paths $C_i$ depicted in FIG.(\ref{zl4}) are followed for the Schwarzschild-AdS black hole in CKG background . Zero points are labeled with black dots, the winding numbers of $\Phi_{11}$ is -1, whereas those of $\Phi_{12}$ is 1.}
\label{zl6}	
\end{center}
\end{figure}
Our analysis reveals that for the CKG parameter, when $\lambda=0.1$, as illustrated in FIG.(\ref{zl1}), there is both a generation point $(\beta_{c1}/ r_0=2.4169)$ and an annihilation point $(\beta_{c2}/ r_0=3.9962)$ present. In contrast, when $\lambda=-0.1$, only an annihilation point $(\beta_{c}/ r_0=3.9332)$ is present, as depicted in FIG.(\ref{zl2}). The vector field, corresponding to the Schwarzschild-AdS black hole within the CKG framework, is illustrated in FIG.(\ref{zl3}) and FIG.(\ref{zl4}). Additionally, the asymptotic characteristics of this field near the boundaries $I_i$ are summarized in TABLE(\ref{table3}). On the one hand, for $\lambda>0$, when $r_h$ approaches infinity, the vector field at boundary $I_1$ shows an orientation directed toward the left. In contrast, as $r_h$ tends to $r_m$, the vector field at boundary $I_3$ displays a direction pointing to the right, which reflects its universal thermodynamic categorization. On the other hand, when $\lambda <0$, it is noted that when $r_h$ gets closer to $r_m$, or as $r$ goes towards infinity,  $\phi$ displays a direction pointing to the right. This suggests that it belongs to a universal thermodynamic classification. Variations in the vector components ($\phi^{r_h}, \phi^{\Theta}$) are systematically examined along the contours depicted in FIGS.(\ref{zl3}) and (\ref{zl4}), with corresponding results presented in FIGS.(\ref{zl5}) and (\ref{zl6}). The zero points located at the origin give rise to closed contour $\Phi_i$ in the vector space when mapped along each $C_i$ path. When $\lambda = 0.1$, it is evident that the winding numbers associated with the zero points are $-1$, $1$, and $-1$ in order, as depicted in FIG.(\ref{zl5}). As a result, both the larger and smaller black holes exhibit instability, whereas the intermediate black hole remains in a stable condition. Their respective heat capacities are less than zero for the unstable black holes and greater than zero for the stable one. As $\beta$ approaches infinity, corresponding to the low-temperature region, the system contains large black holes that display thermodynamic instability as a result of their negative topological charge. In the opposite scenario, as $\beta \rightarrow 0$, which corresponds to high-temperature limits, small black holes emerge. These black holes exhibit similar thermodynamic instability. Therefore, the Schwarzschild-AdS black hole in the CKG background with $(\lambda>0)$ falls into the $W^{1-}$ category.

On the other hand, for $\lambda = -0.1$, two separate black hole states emerge. In terms of winding numbers, small unstable black holes exhibit a value of $-1$, whereas their stable large counterparts are associated with $1$. We then describe the stability of black holes when subjected to constraints of both low and high temperatures. When the temperature is low ($\beta \to \infty$), black hole states are absent. In contrast, at high temperatures ($\beta \to 0$), the system exhibits both unstable small black holes and stable large ones.  As a result, the Schwarzschild-AdS black hole within the CKG framework ($\lambda <0$) is classified into a specific topological category, labeled as $W^{0-}$. When the effect of the CKG parameter ($\lambda = 0$) is disregarded, it becomes clear that the Schwarzschild-AdS black hole remains in the $W^{0-}$ class, which is consistent with the results presented in \cite{gx1}. Furthermore, through the application of a similar analytical methodology, it is revealed that the pressure does not change the topological classification of black holes. More precisely, it is established that the Reissner-Nordstr$\ddot{o}$m black hole, when analyzed within the CKG framework, belongs to the $W^{0+}$ $(\lambda>0)$ and $W^{1+}$ $(\lambda<0)$ classes. In contrast, the Schwarzschild black hole falls into the $W^{1-}$ $(\lambda>0)$ and $W^{0-}$ $(\lambda<0)$ classifications when assessed under the CKG background.

\section{Conclusions}\label{IIII}
\label{Conc}
In conclusion, this work investigates the universal thermodynamic topological classes of static black holes in the CKG background. The paper presents the topological classification and thermodynamic stability of black holes across both low and high temperature ranges, as detailed in TABLE(\ref{tablet}). Our investigation yields several findings, which can be summarized as follows:\\
$(i)$ Our findings indicate that varying the CKG parameter generates two classes of charged AdS black holes: $W^{0+}$ ($\lambda>0$) and $W^{1+}$ ($\lambda<0$). When $\lambda>0$, the low-temperature regime ($\beta \to \infty$) admits stable small black hole solutions, but no black hole states are present at high temperatures ($\beta \to 0$). Conversely, for $\lambda<0$, stable small black holes emerge at low temperature regime ($\beta \to \infty$), while stable large black hole solutions appear at high temperature regime ($\beta \to 0$).  In addition,  the Reissner-Nordstr$\ddot{o}$m black hole, when examined within the context of the CKG framework, is classified as belonging to the $W^{0+}$ $(\lambda > 0)$ and $W^{1+}$ $(\lambda < 0)$ classes.\\
$(ii)$ The CKG parameter is essential in determining the topological classes  of Schwarzschild black holes. In particular, we consider the Schwarzschild-AdS black hole within the CKG framework with $(\lambda>0)$, it is categorized as $W^{1-}$. This classification is marked by the presence of the unstable small and large black holes in both low-temperature and high-temperature conditions, indicating that the CKG alters the outcomes \cite{gx1}. On the other hand, for the CKG parameter set to $(\lambda<0)$, no black hole states exist in the low-temperature regime. However, in the high-temperature regime, two distinct black hole states arise, a stable small black hole and an unstable large one. As a result, the Schwarzschild-AdS black hole in the CKG background with $(\lambda<0)$ is classified as $W^{0-}$. Moreover, when analyzed within the CKG framework, the Schwarzschild black hole falls into the $W^{1-}$ $(\lambda>0)$ and $W^{0-}$ $(\lambda<0)$ classifications.

\acknowledgments
The authors would like to thank our referee for his/her constructive comments and suggestions.
This work is supported by the Doctoral Foundation of Zunyi Normal University of China (BS [2022] 07), by the National Natural Science Foundation of China (NSFC) under Grants No. 12265007, No. 12205243 and No. 12375053. The research of S. Z. and H. H. was supported by the Q-CAYLE project, funded by the European Union-Next Generation UE/MICIU/Plan de Recuperacion, Transformacion y Resiliencia/Junta de Castilla y Leon (PRTRC17.11), and also by project PID2023-148409NB-I00, funded by MICIU/AEI/10.13039/501100011033. Financial support of The Department of Education of the Junta de Castillay Leon and FEDER Funds is also gratefully acknowledged (Reference: CLU-2023-1-05).
Also, H. H. is grateful to Excellence project FoS UHK 2203/2025-2026 for the financial support.
Furthermore, B. C. L. is grateful to Excellence Project P$\check{r}$F UHK  2205 / 2025-2026 for the financial support.
\begin{widetext}
\begin{table*}[ht]
    \centering
    \begin{tabular}{|l|c|c|c|c|c||c|}
        \hline
        \textbf{W}&\textbf{Black hole solutions }& \textbf{Innermost}& \textbf{Outermost} & \textbf{Low $T$} & \textbf{High $T$} & \textbf{DP}  \\
        \hline
        $W^{1-}$& Schwarzschild black hole in CKG $(\lambda>0)$  & \emph{Un} &\emph{Un} & \emph{Un} large & \emph{Un} small & 0  \\\hline
        $W^{0-}$& Schwarzschild black hole in CKG $(\lambda<0)$  &\emph{Un}& \emph{S}& No &Un small+\emph{S} large& $1$ AP \\\hline
        $W^{0+}$& Reissner-Nordstr$\ddot{o}$m black hole in CKG $(\lambda>0)$ & \emph{S} & \emph{Un}& S small+Un large &No & 1GP \\\hline
        $W^{1+}$& Reissner-Nordstr$\ddot{o}$m black hole in CKG $(\lambda<0)$ & \emph{S} & \emph{S}& \emph{S} small &\emph{S} large & 1GP+1AP \\\hline
         $W^{1-}$&Schwarzschild-AdS black hole in CKG $(\lambda>0)$ & \emph{Un} &\emph{Un} & \emph{Un} large & \emph{Un} small & 1GP+1AP \\\hline
        $W^{0-}$&Schwarzschild-AdS black hole in CKG  $(\lambda<0)$  &\emph{Un}& \emph{S}& No &\emph{Un} small+\emph{S} large& $1$ AP \\
        \hline
        $W^{0+}$&Charge-AdS black hole in CKG  $(\lambda>0)$& \emph{S} & \emph{Un}& S small+Un large &No& 2GP+1AP \\\hline
       $W^{1+}$&Charge-AdS black hole in CKG  $(\lambda<0)$ & \emph{S} & \emph{S}& \emph{S} small &\emph{S} large & 1GP+1AP\\
        \hline
    \end{tabular}
    \caption{\emph{Un} and \emph{S} indicate unstable and stable  states, while, GP and AP denote the generation point and annihilation point, respectively.}
   \label{tablet}
\end{table*}
\end{widetext}


\begin{thebibliography}{99}

\def\EPJC{Eur. Phys. J. C\,}
\def\IJMPA{Int. J. Mod. Phys. A\,}
\def\JCAP{J. Cosmol. Astropart. Phys.\,}
\def\JHEP{J. High Energy Phys.\,}
\def\CQG{Classical Quantum Gravity\,}
\def\JMP{J. Math. Phys. (N.Y.)\,}
\def\NPB{Nucl. Phys. B \,}
\def\PDU{Phys. Dark Univ.\,}
\def\PLB{Phys. Lett. B \,}
\def\PRD{Phys. Rev. D\,}
\def\PRL{Phys. Rev. Lett.\,}
\def\PRR{Phys. Rev. Res.\,}
\def\GRG{Gen. Relativ. Gravit.\,}

\bibitem{ch0}
S.W. Hawking,
\href{https://link.springer.com/article/10.1007/s10714-021-02896-y}
{Commun, Math. Phys. \textbf{43 }, 199 (1975)}.
\bibitem{ch01}
S.W. Hawking and D.N. Page,
\href{https://link.springer.com/article/10.1007/BF01208266}
{Commun, Math. Phys. \textbf{87 }, 577 (1983)}.

\bibitem{ch4}
D. Kastor, S. Ray and J. Traschen,
\href{https://doi.org/10.1088/0264-9381/26/19/195011}
{\CQG \textbf{26}, 195011 (2009)}.

\bibitem{ch5}
D. Kastor, S. Ray and J. Traschen,
\href{https://doi.org/10.1088/0264-9381/27/23/235014}
{\CQG \textbf{27}, 235014 (2010)}.

\bibitem{ch6}
D. Kastor, S. Ray and J. Traschen,
\href{https://doi.org/10.1088/0264-9381/28/19/195022}
{\CQG \textbf{28}, 195022 (2011)}.

\bibitem{ch7}
D. Kastor, S. Ray and J. Traschen,
\href{https://doi.org/10.1088/1361-6382/aaf663}
{\CQG \textbf{36}, 024002 (2018)}.
\bibitem{ls0}
S. Chakraborty and T. Padmanabhan,
\href{https://link.springer.com/article/10.1007/JHEP08(2015)029}
{\JHEP \textbf{08},  029 (2015)}.

\bibitem{ls1}
S. Chakraborty and T. Padmanabhan,
\href{https://journals.aps.org/prd/abstract/10.1103/PhysRevD.92.104011}
{\PRD \textbf{92},  104011 (2015)}.

\bibitem{myy1}
N. Altamirano, D. Kubiz\v{n}\'ak and R.B. Mann,
\href{https://doi.org/10.1103/PhysRevD.108.064035}
{\PRD \textbf{88}, 101502 (R) (2013)}.

\bibitem{myy2}
A. M. Frassino, D. Kubiz\v{n}\'ak and R.B. Mann,
\href{https://doi.org/10.1007/JHEP09(2014)080}
{\JHEP \textbf{09} (2014) 080}.


\bibitem{myy3}
D.-C. Zou,  R. Yue and M. Zhang,
\href{https://doi.org/10.1140/epjc/s10052-017-4822-9}
{\EPJC \textbf{77}, 256 (2017)}.

\bibitem{myy4}
S.-W. Wei and Y.-X. Liu,
\href{https://doi.org/10.1103/PhysRevD.90.044057}
{\PRD \textbf{90}, 044057 (2014)}.

\bibitem{myy5}
A. Dehyadegari, A. Sheykhi and A. Montakhab,
\href{https://doi.org/10.1103/PhysRevD.96.084012}
{\PRD 96, 084012 (2017)}.


\bibitem{myy6}
Z. Dayyani and A. Sheykhi,
\href{https://doi.org/10.1103/PhysRevD.98.104026}
{\PRD \textbf{98}, 104026 (2018)}.

\bibitem{myy7}
H. Yazdikarimi, A. Sheykhi and Z. Dayyani,
\href{https://doi.org/10.1103/PhysRevD.99.124017}
{\PRD \textbf{99}, 124017 (2019)}.

\bibitem{myy8}
A. Sheykhi , M. Arab and Z. Dayyani,
\href{https://doi.org/10.1103/PhysRevD.101.064019}
{\PRD \textbf{101}, 064019 (2020)}.

\bibitem{myy9}
S.H. Hendi and K. Jafarzade,
\href{https://doi.org/10.1103/PhysRevD.103.104011}
{\PRD \textbf{103}, 104011 (2021)}.

\bibitem{myy10}
R.H. Ali, G. Abbas and G. Mustafa,
\href{https://doi.org/10.1016/j.dark.2024.101465}
{Phys. Dark Univ. \textbf{44}, 101465 (2024)}.

\bibitem{jyy1}
J. Yang and R.B. Mann,
\href{https://link.springer.com/article/10.1007/JHEP08(2023)028}
{\JHEP \textbf{08}, 028 (2023)}.

\bibitem{jyy2}
R. Li, K. Zhang, J. Yang, R.B. Mann and J. Wang,
\href{https://arxiv.org/abs/2505.24148}
{arXiv: 2505.24148}.
\bibitem{myy11}
\"O. \"Okc\"u and E. Aydiner,
\href{https://doi.org/10.1140/epjc/s10052-017-4598-y}
{\EPJC \textbf{77}, 24 (2017)}.

\bibitem{myy12}
\"O. \"Okc\"u and E. Aydiner,
\href{https://doi.org/10.1140/epjc/s10052-018-5602-x}
{\EPJC \textbf{78}, 123 (2018)}.

\bibitem{myy16}
M.-Y. Zhang, H. Chen, H. Hassanabadi, Z.-W. Long and H. Yang,
\href{https://doi.org/10.1088/1674-1137/aca958}
{Chin. Phys. C \textbf{47}, 045101 (2023)}.
\bibitem{myy18}
M.-Y. Zhang,  H. Chen,  H. Hassanabadi, Z.-W. Long and H. Yang,
\href{https://doi.org/10.1088/1674-1137/ad32c0}
{Chin. Phys. C \textbf{48}, 065101 (2024)}.

\bibitem{myy13}
S.-Q. Lan,
\href{https://doi.org/10.1103/PhysRevD.98.084014}
{\PRD \textbf{98}, 084014 (2018)}.

\bibitem{myy14}
S. Bi, M. Du, J. Tao and F. Yao,
\href{https://doi.org/10.1088/1674-1137/abcf23}
{Chin. Phys. C \textbf{45}, 025109 (2021)}.

\bibitem{myy15}
J. Liang, B. Mu and P. Wang,
\href{https://doi.org/10.1103/PhysRevD.104.124003}
{\PRD \textbf{104}, 124003 (2021)}.

\bibitem{myy17}
M. Yasir, X. Tiecheng, F. Javed and G. Mustafa,
\href{https://doi.org/10.1088/1674-1137/ad0962}
{Chin. Phys. C \textbf{48}, 015103 (2024)}.

\bibitem{G2}
S.-W. Wei and Y.-X. Liu,
\href{https://journals.aps.org/prd/abstract/10.1103/PhysRevD.105.104003}
{\PRD \textbf{105}, 104003 (2022)}.
\bibitem{G1}
 S.W. Wei, Y.X. Liu and R.B. Mann,
 \href{https://journals.aps.org/prl/abstract/10.1103/PhysRevLett.129.191101}
{\PRL \textbf{129}, 191101 (2022)}.

\bibitem{ll1}
S.-W. Wei and Y.-X. Liu,
\href{https://doi.org/10.1103/PhysRevD.107.064006}
{\PRD \textbf{107}, 064006 (2023)}.
\bibitem{gx2}
S.-W. Wang,S.-P. Wu and S.-W. Wei,
\href{https://www.sciencedirect.com/science/article/pii/S0370269325001625? }
{\PLB \textbf{864}, 139402 (2025)}.
\bibitem{ll2}
X. Ye and S.-W. Wei,
\href{https://doi.org/10.1088/1475-7516/2023/07/049}
{\JCAP \textbf{07} (2023) 049}.

\bibitem{ll3}
D. Wu,
\href{https://doi.org/10.1103/PhysRevD.107.024024}
{\PRD \textbf{107}, 024024 (2023)}.

\bibitem{ll4}
D. Wu and S.-Q. Wu,
\href{https://doi.org/10.1103/PhysRevD.107.084002}
{\PRD \textbf{107}, 084002 (2023)}.

\bibitem{ll5}
D. Wu,
\href{https://doi.org/10.1140/epjc/s10052-023-11561-4}
{\EPJC \textbf{83}, 365 (2023)}.

\bibitem{ll6}
D. Wu,
\href{https://doi.org/10.1140/epjc/s10052-023-11782-7}
{\EPJC \textbf{83}, 589 (2023)}.

\bibitem{ll7}
D. Wu,
\href{https://doi.org/10.1103/PhysRevD.108.084041}
{\PRD \textbf{108}, 084041 (2023)}.

\bibitem{ll8}
D. Wu, S.-Y. Gu, X.-D. Zhu, Q.-Q. Jiang and S.-Z. Yang,
\href{https://doi.org/10.1007/JHEP06(2024)213}
{\JHEP \textbf{06} (2024) 213}.

\bibitem{ll9}
X.-D. Zhu, D. Wu and D. Wen,
\href{https://linkinghub.elsevier.com/retrieve/pii/S0370269324004775}
{\PLB \textbf{856} (2024) 138919}.

\bibitem{ll34}
W. Liu, L. Zhang, D. Wu and J. Wang,
\href{https://arxiv.org/abs/2409.11666}
{arXiv: 2409.11666}.

\bibitem{ll10}
P.K. Yerra and C. Bhamidipati,
\href{https://doi.org/10.1103/PhysRevD.105.104053}
{\PRD \textbf{105}, 104053 (2022)}.

\bibitem{ll11}
P.K. Yerra and C. Bhamidipati,
\href{https://doi.org/10.1016/j.physletb.2022.137591}
{\PLB \textbf{835}, 137591 (2022)}.

\bibitem{ll12}
P.K. Yerra, C. Bhamidipati and S. Mukherji,
\href{https://doi.org/10.1103/PhysRevD.106.064059}
{\PRD \textbf{106}, 064059 (2022)}.

\bibitem{ll13}
P.K. Yerra, C. Bhamidipati and S. Mukherji,
\href{https://doi.org/10.1088/1742-6596/2667/1/012031}
{J. Phys. Conf. Ser. \textbf{2667}, 012031 (2023)}.
\bibitem{ll14}
P.K. Yerra, C. Bhamidipati and S. Mukherji,
\href{https://doi.org/10.1007/JHEP03(2024)138}
{\JHEP \textbf{03} (2024) 138}.

\bibitem{ll15}
N.-C. Bai, L. Li and J. Tao,
\href{https://doi.org/10.1103/PhysRevD.107.064015}
{\PRD \textbf{107}, 064015 (2023)}.

\bibitem{ll16}
N.-C. Bai, L. Song, and J. Tao,
\href{https://doi.org/10.1140/epjc/s10052-024-12407-3}
{\EPJC \textbf{84}, 43 (2024)}.

\bibitem{ll17}
M.-Y. Zhang, H. Chen, H. Hassanabadi, Z.-W. Long and H. Yang,
\href{https://doi.org/10.1140/epjc/s10052-023-11933-w}
{\EPJC \textbf{83}, 773 (2023)}.
\bibitem{ll18}
M.-Y. Zhang, H. Chen, H. Hassanabadi, Z.-W. Long and H. Yang,
\href{https://www.sciencedirect.com/science/article/pii/S037026932400443X?via}
{\PLB \textbf{856}, 138885 (2024)}.


\bibitem{ll19}
M.-Y. Zhang,H.-Y. Zhou, H. Chen, H. Hassanabadi and Z.-W. Long
\href{https://link.springer.com/article/10.1140/epjc/s10052-024-13586-9}
{\EPJC \textbf{84}, 1251 (2024)}.

\bibitem{ll20}
H. Chen,D. Wu, M.-Y. Zhang, H. Hassanabadi and Z.-W. Long
\href{https://www.sciencedirect.com/science/article/abs/pii/S2212686424001997?via}
{\PDU \textbf{46}, 101617 (2024)}.
\bibitem{ll21}
M.R. Alipour, M.A.S. Afshar, S.N. Gashti, and J. Sadeghi,
\href{https://doi.org/10.1016/j.dark.2023.101361}
{\PDU \textbf{42}, 101361 (2023)}.

\bibitem{ll22}
J. Sadeghi, S.N. Gashti, M.R. Alipour, and M.A.S. Afshar,
\href{https://doi.org/10.1016/j.aop.2023.169391}
{Ann. Phys. (Amsterdam) \textbf{455}, 169391 (2023)}.

\bibitem{ll23}
J. Sadeghi, M.A.S. Afshar, S.N. Gashti and M.R. Alipour,
\href{https://doi.org/10.1016/j.aop.2023.169569}
{Ann. Phys. (Amsterdam) \textbf{460} , 169569 (2023)}.

\bibitem{ll24}
J. Sadeghi, M.A.S. Afshar, S.N. Gashti and M.R. Alipour,
\href{https://doi.org/10.1088/1402-4896/ad186b}
{Phys. Scripta \textbf{99}, 025003 (2024)}.
\bibitem{ll25}
N.J. Gogoi and P. Phukon,
\href{https://doi.org/10.1103/PhysRevD.108.066016}
{\PRD \textbf{108}, 066016 (2023)}.

\bibitem{ll26}
N.J. Gogoi and P. Phukon,
\href{https://doi.org/10.1016/j.dark.2024.101456}
{\PDU \textbf{44}, 101456 (2024)}.

\bibitem{ll27}
N.J. Gogoi and P. Phukon,
\href{https://doi.org/10.1103/PhysRevD.107.106009}
{\PRD \textbf{107}, 106009 (2023)}.

\bibitem{yin1}
P. Yin, Y. Liu, J. Chen and Y. Wang,
\href{https://inspirehep.net/files/ff8ad31336a84d371cb65e00c558907b}
{\EPJC \textbf{ 85}, 661 (2025)}.

\bibitem{ll28}
M.S. Ali, H.E. Moumni, J. Khalloufi and K. Masmar,
\href{https://doi.org/10.1016/j.aop.2024.169679}
{Annals Phys. \textbf{465}, 169679 (2024)}.

\bibitem{ll29}
K. Bhattacharya, K. Bamba and D. Singleton,
\href{https://doi.org/10.1016/j.physletb.2024.138722}
{\PLB \textbf{854}, 138722 (2024)}.

\bibitem{ll30}
Z.-Y. Fan,
\href{https://doi.org/10.1103/PhysRevD.107.044026}
{\PRD \textbf{107}, 044026 (2023)}.

\bibitem{ll31}
C.X. Fang, J. Jiang and M. Zhang,
\href{http://dx.doi.org/10.1007/JHEP01(2023)102}
{\JHEP \textbf{01} (2023) 102}.


\bibitem{ll32}
M. Zhang and J. Jiang,
\href{https://doi.org/10.1007/JHEP06(2023)115}
{\JHEP \textbf{06} (2023) 115}.

\bibitem{ll33}
Y. Du and X. Zhang,
\href{https://doi.org/10.1140/epjc/s10052-023-12114-5}
{\EPJC \textbf{83}, 927 (2023)}.
\bibitem{gx1}
S.-W. Wei, Y.-X. Liu and Robert B. Mann,
\href{https://journals.aps.org/prd/abstract/10.1103/PhysRevD.110.L081501}
{\PRD \textbf{110}, L081501 (2024)}.
\bibitem{kh6}
Y.S. Duan,
The structure of the topological current,
SLAC-PUB-3301, (1984).

\bibitem{gx4}
D. Wu,  W. Liu, S.-Q. Wu and Robert B. Mann,
\href{https://journals.aps.org/prd/abstract/10.1103/PhysRevD.111.L061501}
{\PRD \textbf{111}, L061501 (2025)}.


\bibitem{gx3}
Y. Chen, X.-D. Zhu and D. Wu,
\href{https://www.sciencedirect.com/science/article/pii/S0370269325002436?via} {\PLB \textbf{865}, 139482 (2025)}

\bibitem{gx5}
X.-D. Zhu, W. Liu and D. Wu,
\href{https://www.sciencedirect.com/science/article/pii/S0370269324007214?}
{\PLB \textbf{860}, 139163 (2025)}.

\bibitem{gx6}
M. Rizwan, M. Jamil and M.Z. A. Moughal,
\href{https://www.sciencedirect.com/science/article/pii/S0370269324007214}
{\EPJC \textbf{85}, 359 (2025)}.
\bibitem{my1}
B. P. Abbott et. al,
\href{https://link.springer.com/article/10.1007/s10714-021-02896-y}
{\PRL \textbf{116 },131103 (2016)}.
\bibitem{my2}
B. P. Abbott et. al,
\href{https://journals.aps.org/prd/abstract/10.1103/PhysRevD.93.122003}
{\PRD \textbf{93}, 122003 (2016)}.
\bibitem{my3}
K. Akiyama et. al,
\href{https://iopscience.iop.org/article/10.3847/2041-8213/ab0ec7}
{Astrophys. J. Lett. \textbf{875}, L1 (2019)}.
\bibitem{my4}
K. Akiyama et. al,
\href{https://iopscience.iop.org/article/10.3847/2041-8213/ab0c96}
{Astrophys. J. Lett. \textbf{875}, L2 (2019)}.
\bibitem{my5}
K. Akiyama et. al,
\href{https://iopscience.iop.org/article/10.3847/2041-8213/ab0c57}
{Astrophys. J. Lett. \textbf{875}, L3 (2019)}.
\bibitem{my6}
K. Akiyama et. al,
\href{https://iopscience.iop.org/article/10.3847/2041-8213/ab0e85}
{Astrophys. J. Lett. \textbf{875}, L4 (2019)}.

\bibitem{my7}
K. Akiyama et. al,
\href{https://iopscience.iop.org/article/10.3847/2041-8213/ac6674}
{Astrophys. J. Lett. \textbf{930}, L12 (2022)}.
\bibitem{my8}
K. Akiyama et. al,
\href{https://iopscience.iop.org/article/10.3847/2041-8213/ac6429}
{Astrophys. J. Lett. \textbf{930}, L14 (2022)}.

\bibitem{my9}
T. Clifton, P. G. Ferreira, A. Padilla and C. Skordis,
\href{https://www.sciencedirect.com/science/article/pii/S0370157312000105?via}
{Phys. Rept. \textbf{513}, 189 (2012)}.
\bibitem{my10}
P. Bull, et al,
\href{https://www.sciencedirect.com/science/article/abs/pii/S2212686416300097?}
{Phys. Dark Univ. \textbf{12}, 56 (2016)}.

\bibitem{my11}
G. Bertone, D. Hooper and J. Silk,
\href{https://www.sciencedirect.com/science/article/abs/pii/S0370157304003515?}
{Phys. Rept. \textbf{405}, 279 (2005)}.
\bibitem{my12}
P. J. E. Peebles and B. Ratra,
\href{https://journals.aps.org/rmp/abstract/10.1103/RevModPhys.75.559}
{Rev. Mod. Phys. \textbf{75}, 559 (2003)}.
\bibitem{my13}
S. Capozziello and M. De Laurentis,
\href{https://www.sciencedirect.com/science/article/abs/pii/S0370157311002432?}
{Phys. Rept. \textbf{509}, 167 (2011)}.
\bibitem{my14}
C. A. Mantica and L. G. Molinari,
\href{https://journals.aps.org/prd/abstract/10.1103/PhysRevD.108.124029}
{\PRD \textbf{108}, 124029 (2023)}.
\bibitem{my15}
C. A. Mantica and L. G. Molinari,
\href{https://journals.aps.org/prd/abstract/10.1103/PhysRevD.110.044025}
{\PRD \textbf{110}, 044025 (2024)}.

\bibitem{mm2.1}
J. Harada,
\href{https://journals.aps.org/prd/abstract/10.1103/PhysRevD.104.088502}
{\PRD \textbf{104}, 088502 (2021)}.
\bibitem{mm2}
J. Harada,
\href{https://journals.aps.org/prd/abstract/10.1103/PhysRevD.103.L121502}
{\PRD \textbf{103}, L121502 (2021)}.
\bibitem{mm1}
J. Harada,
\href{https://journals.aps.org/prd/abstract/10.1103/PhysRevD.106.064044}
{\PRD \textbf{106}, 064044 (2022)}.

\bibitem{mm3}
J. Harada,
\href{https://journals.aps.org/prd/abstract/10.1103/PhysRevD.108.044031}
{\PRD \textbf{108}, 044031 (2023)}.
\bibitem{mm3.1}
J. Harada,
\href{https://journals.aps.org/prd/abstract/10.1103/PhysRevD.108.104037}
{\PRD \textbf{108}, 104037 (2023)}.

\bibitem{mm4}
P. Xia, D. Zhang, X. Ren, B. Wang and Y. C. Ong,
\href{https://journals.aps.org/prd/abstract/10.1103/PhysRevD.111.023526}
{\PRD \textbf{111}, 023526 (2025)}.

\bibitem{HaradaPRD2023}
J. Harada,
\href{https://doi.org/10.1103/PhysRevD.108.044031}
{\PRD \textbf{108}, 044031 (2023)}.


\bibitem{mm5}
J. T. S. S. Junior, F. S. N. Lobo and M. E. Rodrigues,
\href{https://iopscience.iop.org/article/10.1088/1361-6382/ad210e}
{\CQG \textbf{41}, 055012 (2024)}.
\bibitem{mk1}
S. Ghaffari and G. G. Luciano,
\href{https://arxiv.org/abs/2505.06560}
{arXiv: 2505.06560}.

\bibitem{mm8}
D.~Kubiznak and R.~B.~Mann,
\href{https://link.springer.com/article/10.1007/JHEP08(2022)1743}
{\JHEP \textbf{07}, 033 (2012)}.

\bibitem{mm9}
G.~G.~Luciano and A.~Sheykhi,
\href{https://www.sciencedirect.com/science/article/pii/S221268642300153X?}
{Phys. Dark Univ. \textbf{42}, 101319 (2023)}.
\end{thebibliography}
\end{document}